\documentclass[prl,twocolumn,showpacs,preprintnumbers,amsmath,amssymb]{revtex4}

\usepackage{epsfig}
\usepackage{dcolumn}
 \usepackage{bm}

\begin{document}

\preprint{
\hfill  TH01.1
} 

\title{Hidden Symmetries in Nano-magnets}

\author{M. Preda and S. E. Barnes }
\affiliation{Department of Physics, University of Miami, Coral Gables, 
Florida 33124}

\date{\today} 
\begin{abstract}
{The hidden symmetry of certain nano-magnets leads to many of the 
levels being doubly degenerate for periodic values of the Zeeman 
energy.  Corresponding to such a symmetry is an operator, $K_{n}$, 
related to the time reversal operator, and which commutes with the 
Hamiltonian.  The degeneracies for whole integer spin values are often 
similar to the Kramer's degeneracy for half-integer spin.  An all 
algebraic method for constructing Hamiltonians with such hidden 
symmetries and different crystal symmetries is described.  }
\end{abstract}


\maketitle

For both of the much studied molecular magnets 
Mn${}_{12}$\cite{thomas} and Fe${}_{8}$\cite{ws} it is found 
experimentally that most of the levels cross, thereby becoming doubly 
degenerate, at a set of periodic values of the magnetic fields applied 
along a certain symmetry direction.  In fact, neither of these 
statements is {\it precisely\/} true.  Levels cross at {\it almost\/} 
the same field and the fields at which this occurs are {\it almost\/} 
periodic up to some maximum value.  Theoretically this phenomena of 
simultaneous degeneracies has been described in terms of topological 
interference at ``diabolic'' points\cite{garg} and in terms of 
intermediate spin\cite{barnes}.  In both these theoretical approaches 
certain approximations need to be made.

In this Letter it is shown that the degeneracy of nearly all the 
levels reflects a hidden symmetry, i.e., that there exists an 
operator, $K_{n}$, related to the time reversal operator which 
commutes with the Hamiltonian for the special values of the field 
$B_{n}$.  It will thereby be shown how to construct spin models which 
have the property that many of the levels cross at {\it exactly\/} the 
same field for {\it exactly\/} periodic fields (up to some maximum), 
this independent of the spin value $S$ and even if the parameters 
correspond to the tunneling regime.  In effect, it is shown that {\it 
whole\/} integer spins, in a suitable finite magnetic field $B_{1}$, 
exhibit a Kramer's degeneracy, for all but a single highest energy 
level.  This supports the intermediate spin 
interpretation\cite{barnes} of the situation since in effect a whole 
integer spin has been converted into one which for practical purposes 
behaves as if it was half integer.  Unlike other approaches to this 
problem the present one is purely algebraic and without 
approximations.  (Although the same algebraic methods will be used to 
generate useful approximate results.)  The result has implications for 
magnetization plateaus for concentrated bulk magnets.

Experimentally relaxation via tunneling is exactly suppressed at 
applied fields for which there is such a hidden symmetry. This has 
implications for applications in both magnetic recording and quantum 
computing.

The large spin model appropriate to Fe${}_{8}$ is traditionally 
written\cite{ws} as,
\begin{equation}
{\cal H} = - D {S_{y}}^{2} + E[{S_{x}}^{2}-{S_{y}}^{2}] 
+ g \mu_{B} B S_{x}
\label{un}
\end{equation}
where without loss of generality the positive parameters have $D> 
E$\cite{preda} and where the coordinate system coincides with the 
principal crystal axes. 

The first exercise is to show for this model all the levels but one 
exhibit a Kramer's degeneracy when $B = \hbar \sqrt{2E(D+E)}$.  The 
existence of this massive degeneracy is intimately related to the 
structure of the SU(2) Lie algebra.  Consider the most general {\it 
linear\/} combination of the elements of this algebra, written as,
\begin{equation}
r = \alpha S_{x} + i \beta S_{y},
\label{deux}
\end{equation}
where $S_{z}$ has been eliminated by a suitable choice of axes. Take 
$T$ to be the time reversal operator for integer spin, so $T^{2} = 1$, 
then,
\begin{equation}
r^{\dagger} = -T r T,
\label{trois}
\end{equation}
independent of the $c$-numbers $\alpha$ and $\beta$.  The 
pseudo-time-reversal-operator is, written as,
\begin{equation}
K_{1} = T r.
\label{quatre}
\end{equation} 
This has the evident property that $[{\cal H}, K_{1}] = 0$, if,
\begin{equation}
{\cal H}_{1} =  {K_{1}}^{2} = - r^{\dagger} r.
\label{cinq}
\end{equation}
Using the definition of $r$ with $\alpha = \sqrt{D+E}$ and $\beta = 
\sqrt{2E}$, 
\begin{equation}
{\cal H}_{1} = - (D + E)  {S_{x}}^{2} -2 E{S_{y}}^{2} 
+  \hbar \sqrt{2E(D+E)} S_{z},
\label{six}
\end{equation}
which, to within a constant, is equivalent to Eqn.~(\ref{un}) with $g 
\mu_{B} B = \hbar \sqrt{2E(D+E)}$ and with a change of axis so that 
$x$ is the easy direction.  If $|n\rangle$ is an eigenstate of ${\cal 
H}_{1}$ with energy $E_{n}$ then $K_{1} |n\rangle$ is either (i) zero 
or (ii) another degenerate eigenstate of ${\cal H}_{1}$.  In the 
standard way, ${\cal H}_{1} K_{1} |n\rangle = K_{1} {\cal H}_{1} 
|n\rangle = E_{n }K_{1} |n\rangle$, then for the normalization,
\begin{equation}
\left \langle n \right|r ^{\dagger} T^{\dagger} Tr \left| n \right\rangle
=
\left \langle n \right|r ^{\dagger} r \left| n \right\rangle
= - E_{n},
\label{sixbis}
\end{equation}
since $T$ is Hermitian, i.e., $T^{\dagger} T = 1$ and using ${\cal 
H}_{1} = - r ^{\dagger} r$.  Thus (i) if $K_{1}$ destroys $|n\rangle$ 
then $E_{n} = 0$ and otherwise (ii) when the norm of $K_{1} |n\rangle$ 
is finite then $E_{n}< 0$.  It is necessary to show that $K_{1} 
|n\rangle$ and $|n\rangle$ are different states.  To this end it is 
useful to write:
\begin{eqnarray}
{\cal H}_{1} &=& - r^{\dagger} r \nonumber \\
& =& - 
(\alpha^{2} - \beta^{2} )({S^{+}}^{2} + {S^{-}}^{2} ) \nonumber\\
&&\ \ \  - (\alpha + 
\beta )^{2} S^{-} S^{+} - (\alpha - \beta )^{2} S^{+} S^{-},
\label{xyz}
\end{eqnarray}
which makes evident the fact that $\cal H$ breaks into two blocks.  It 
is possible to insist that there exists a normalized $\left|n, e 
\right\rangle = \sum_{m}a_{2m}\left|S, 2m \right\rangle$ which 
involves only the eigenstates of $S_{z}$ with {\it even\/} eigenvalues 
$2m$, i.e, $S_{z}\left|S, 2m \right\rangle = 2m\hbar \left|S, 2m 
\right\rangle$.  Similarly there is the normalized odd state $\left| 
n ,o \right\rangle = \sum_{m}a_{2m+1}\left|S, 2m+1 \right\rangle$.  It 
is then observed that $K_{1} = T r = T [(\alpha + \beta ) S^{-} + 
(\alpha - \beta ) S^{+}]$, so that $K_{1}\left| n,e \right\rangle 
\propto \left| n,o \right\rangle$ and $\left| n, e \right\rangle$ and 
$\left|n,o \right\rangle$ {\it are\/} independent.

Since for whole integer $S$, the total number of states, $2S+1$, is 
odd there must be {\it at least\/} one state which is not part of a 
doublet and therefore for which $E_{n} =0$.  That this state is unique 
is proved by explicit solution of the equation $K_{1} \left| n 
\right\rangle = 0$.  The resulting matrix equation is,
\begin{eqnarray}
&& \left( 
\begin{array}{ccccc} 
0  & \cos \alpha M_{S}^{S-1}&0& \ldots&0 \\
\sin \alpha   &0& \cos \alpha  M_{S-1}^{S-2}& \ldots &0\\
0  &\sin \alpha M_{S-1}^{S-2} &      0& \ldots & 0 \\
\vdots        & \vdots     & \vdots        &\ddots & \vdots \\
 0            & 0            &     0 & \ldots &0       
 \end{array}
 \right)
\left(\begin{array}{c}
a_{S} \\ a_{S-1}  \\ a_{S-2} \\ \vdots\\ a_{-S} 
\end{array}\right)
 =
 \nonumber \\
&& \left( 
\begin{array}{c}
 a_{S-1} \cos \alpha M_{S}^{S-1} \hfill \\
 a_{S} \sin \alpha  M_{S}^{S-1} + a_{S-2} \cos \alpha M_{S-1}^{S-2}\\
 a_{S-1} \sin \alpha M_{S-1}^{S-2} + a_{S-3} \cos \alpha 
 M_{S-2}^{S-3} \\
 \vdots \\
 a_{-S+1} \sin \alpha M_{-S}^{-S+1} \hfill \end{array}\right)
  = 0.
\label{sept}
\end{eqnarray}
where $M_{n}^{m} = [S(S+1) - nm]^{1/2}$.  This reduces to:
\begin{eqnarray}
&& a_{S-1}=a_{S-3}=\ldots=a_{-S+1}=0 \qquad \hbox{  and  }
\nonumber \\
&&
 a_{S-2}=-\frac{\sin \alpha M_{S}^{S-1} }{ \cos \alpha 
 M_{S-1}^{S-2}}a_{S},\quad 
 a_{S-4}= -\frac{\sin \alpha M_{S-2}^{S-3}  }{\cos \alpha 
 M_{S-3}^{S-4} } a_{S-2}, 
\nonumber \\
&&
 \ldots a_{-S}=-\frac{\sin \alpha  M_{-S+2} }{ \cos \alpha  
 M_{-S+1}^{-S}}a_{-S+2}.
\label{huit}
\end{eqnarray}
The solution is uniquely determined up to a multiplicative constant, 
proving the uniqueness of the state with $E_{n} = 0$.

\begin{figure}[t!]
\centerline{\epsfig{file=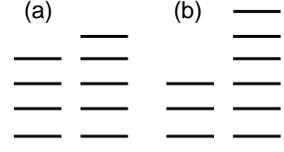,width=1.4in}  } 
\vspace{-3pt} \caption[toto]{(a) Illustrated for $S=4$ is the 
situation for the first magic field $g \mu_{B} B_{1} = \hbar 
\sqrt{2E(D+E)}$.  There are four pseudo-Kramer's doublets with a 
highest energy singlet.  The singlet is at zero energy.  (b) At the 
second magic field $B_{2}$ for $S=4$, there are three doublets with 
three singlets at higher energy.  }
\label{fig1}
\end{figure}

The situation is illustrated in Fig.~\ref{fig1}a, all the lowest levels 
form pseudo-Kramer's doublets with the sole exception of the highest 
energy state which has zero energy.  If one ignores the highest energy 
singlet, the spectrum is identical to that for half-integer spin $S = S 
-(1/2)$.  These conclusions are independent of the parameters $D$ and 
$E$ and the value of $S$, i.e., these precise degeneracies occur even 
in the absence of a tunneling barrier.

It is the case that $[r,r^{\dagger}] = 2 \hbar \sqrt{2E(D+E)}S_{z}$ 
and so the second Hamiltonian in the series is,
$$
{\cal H}_{2} = - r^{\dagger}r + [r,r^{\dagger}],
$$
which corresponds to $g \mu_{B} B_{2} = 3 \hbar \sqrt{2E(D+E)}$.  This 
also has pseudo-Kramer's degeneracies.  To prove this it is necessary 
to find a $K_{2}$ such that $[{\cal H}_{2},K_{2}] =0$.  Since $r$ is 
related to the raising operator a first guess might be $Tr^{3}$, 
however the relevant commutator is rather,
\begin{equation}
\left[{\cal H}_{2}, T r^{3}\right]= 2 \alpha \beta (\beta^{2}-
\alpha^{2})  \left[T r , S_{z} \right]\neq 0.
	\label{onze}
\end{equation}
The commutator to the right is that involved in $[{\cal H}_{2},K_{1}] 
= -2 \alpha \beta [Tr, S_{z}]$, and it follows that,
\begin{equation}
K_{2}= T r^{3} +( \beta ^{2}- \alpha^{2} ) Tr. 
\label{douze}
\end{equation}

The special case $E=D$, when ${\cal H}_{2} = 2ES_{z}(S_{z}+3)$ and 
$K_{2} = T(S_{+})^{3}$, then establishes that $K_{2}$ destroys three 
states which are singlets, Fig.~\ref{fig1}b.

The generalization is straightforward: ${\cal H}_{n} = - r^{\dagger}r + 
(n-1)[r,r^{\dagger}]$ and $K_{n}$ has the form:
\begin{equation}
K_{n}= \sum_{k=0}^{n-1} a_{n}^{k} (\beta^{2}-\alpha^{2})^{n-k-1} Tr^{2k+1} 
\label{treize}
\end{equation}
where $ a_{n}^{k}$ are integers, and where it is possible to set 
$a_{n}^{n-1}=1$.  The power of $(\beta^{2}- \alpha^{2})$ ensures that 
all terms $(\beta^{2}-\alpha^{2})^{n-k-1} T r^{2k+1}$ have the 
dimension $({\rm Energy})^{n+\frac{1 }{ 2}}$.  

It is an interesting but tedious exercise in linear algebra to 
construct an algorithm to determine the $a_{n}^{k}$.  The results for 
small $n$ are:
\begin{eqnarray}
&& n=3: \quad a_{3}^{1} =5, \quad a_{3}^{0}=4,
\nonumber \\
&& 
n=4: \quad a_{4}^{2}=14, \quad a_{4}^{1}=49,\quad  
a_{4}^{0}=36.
\label{quatorze}
\end{eqnarray}
The method and results for several other values of $n$ are given 
elsewhere\cite{preda}. Again, the special case when $E=D$ illustrates that 
there are $2n-1$ singlets for a field $B_{n} = (2n-1)\hbar 
\sqrt{2E(D+E)}$ so that $n=S$ corresponds to the largest value of 
$B_{n}$ for which there is a pseudo-Kramer's doublet.

In summary, it has been proved that there exist a series of operators 
$K_{n}$ which commute with the $\cal H$ of Eqn.~({\ref{un}) for exactly 
periodic magic values $B_{n}$ of the applied field along hardest axis 
of the nano-magnet. For such values of the field there are $2S+1 - 
2n$ pseudo-Kramer's doublets which are exactly degenerate.

The highest symmetry of Eqn.~({\ref{un}) is two-fold.  It is 
interesting to investigate the problem with a four-fold axis.  The 
simplest possibility is to set,
\begin{equation}
r = \eta {S^{+}}^{2} + \gamma {S^{-}}^{2}
\label{dixhuit}
\end{equation}
whence $r^{\dagger} = TrT$.  As before define $K = Tr$, then ${\cal H}_{1} 
= - K^{2}$ can be written as,
\begin{eqnarray}
&&{\cal H}_{1} = -K^{2} = -(\eta^{2}+\gamma^{2}) 
[{S_{z}}^{4} + (5-2S(S+1)){S_{z}}^{2}]
\nonumber  \\
&&
- \eta\gamma [ {S^{+}}^{4} + {S^{-}}^{4}]
\nonumber  \\
&&
- (\eta^{2}-\gamma^{2}) [4{S_{z}}^{3} - (2-4S(S+1))S_{z}].
\label{dixneuf}
\end{eqnarray}
This $\cal H$ has the desired symmetry.  The diagonal anisotropy term 
has the ratio of the coefficients of ${S_{z}}^{4}$ and ${S_{z}}^{2}$ 
fixed and the Zeeman term contains ${S_{z}}^{3}$ in addition to 
$S_{z}$.  For all values of $S$ this corresponds to an easy plane 
magnet.  Notice that the off-diagonal part $\left({S^{+}}^{4} + 
{S^{-}}^{4}\right)$ connects states with $S_{z}$ quantum numbers which 
differ by four and implies that the matrix $\cal H$ breaks down into 
four blocks.  Consider whole integer spin $S$, the blocks behave 
differently as the values of $S_{z}$ are even or odd.  The operator 
$K$ acting upon a solution, $|n\rangle$ which contains the state with 
$S_{z} =0$ produces a state $K|n\rangle$ in a different block, i.e., 
that which involves the $S_{z} =2$ state.  However when $K$ operates 
on a state which contains either of the $S_{z} = \pm 1$ states, the 
result belongs to the same block and the limit $\eta\gamma \to 0$ 
shows $K|n\rangle$ is {\it not\/} distinct from $|n\rangle$.  This 
should be compared with the situation when the Zeeman term, 
$(\eta^{2}-\gamma^{2}) [4{S_{z}}^{3} - (2-4S(S+1))S_{z}]$, is absent.  
Then the plain time reversal operator $T$ commutes with $\cal H$ and 
connects the $S_{z} = + 1$ to the $S_{z} = - 1$ blocks but now maps 
the solutions in the $S_{z} =0$ and $S_{z} =2$ blocks back to 
themselves.  Thus, e.g., if $2S+1$ is a multiple of four there are 
half of the states which are degenerate {\it both\/} in the absence of 
a Zeeman splitting {\it and\/} for the sequence of Hamiltonians ${\cal 
H}_{n} = {\cal H}_{1}+(n-1) [r,r^{\dagger}]$ and operators $K_{n}$ 
constructed by the present methods, i.e., these ${\cal H}_{n}$ 
correspond to the degeneracies of a {\it whole integer point\/} rather 
that a half integer point for which Kramer's theorem insists that {\it 
all\/} states form doublets.

A second possibility is to take,
\begin{equation}
r = D^{1/2}\left(S^{+} + \frac{a }{ D}{S^{-}}^{3} \right),
\label{dixhuitbis}
\end{equation}
whence $r^{\dagger} = -T rT$ and,
\begin{eqnarray}
&& {\cal H}_{1}=  K^{2} = (Tr)^{2} = - r^{\dagger} r
\nonumber  \\
&&
 = - D  S^{-}S^{+}
- a \left({S^{+}}^{4} + {S^{-}}^{4}\right)
- \frac{a^{2} }{ D}{S^{+}}^{3}{S^{-}}^{3}
\nonumber  \\
&&
 =  D {S_{z}}^{2} - a \left({S^{+}}^{4} + {S^{-}}^{4}\right)
\nonumber  \\
&& \ \ + H + (D + \frac{a^{2}}{D}Z)S_{z},
\label{dixneufbis}
\end{eqnarray}
where $H = - DS(S+1) - (a^{2}/D)[M^{4}-8M^{2}+ M(23{S_{z}}^{2}+12) - 
28 {S_{z}}^{2}$, with $M = S(S+1) - {S_{z}}^{2}$, and where the Zeeman 
term involves,
\begin{eqnarray}
Z({S_{z}}^{2}) = &3&[2S^{2}(S+1)^{2}+12S(S+1)+4]
\nonumber  \\
&-&[18S(S+1) - 51]{S_{z}}^{2} + 9 
{S_{z}}^{4}.
\label{quatorzebis}
\end{eqnarray}
With this choice of $r$ is is possible to change the ratio of 
${S_{z}}^{4}$ and ${S_{z}}^{2}$ in the diagonal anisotropy but at the 
cost of a more complicated Zeeman term, $[r,r^{\dagger}] = (D + 
\frac{a^{2}}{D}Z)S_{z}$.  However both terms simplify in the limit 
that $D \gg a$ whence the magic values have $g \mu_{B} B_{n} = (2n-1) 
\hbar D$.  The operators $K_{n}$ now connect even and odd $S_{z}$ 
blocks and so the ${\cal H}_{n} = {\cal H}_{1}+(n-1) [r,r^{\dagger}]$ 
and $K_{n}$ correspond to the double degeneracies of a {\it half integer 
point\/} as for the development with two fold symmetry.

As an alternative approach to four fold symmetry, consider the ${\cal 
H}_{1}$, of Eqn.~(\ref{six}) with $E=1$ and $D=3$.  This can be 
written as,
\begin{equation}
{\cal H}_{1}  = - r^{\dagger} r = - 2S(S+1) + 2({S_{z}}^{2} - {S_{x}}^{2})
 + 2 \sqrt{2} S_{z}.
\label{quinze}
\end{equation}
The square of ${\cal H}_{1} + 2S(S+1) - 1 = -1 + 2({S_{z}}^{2} - 
{S_{x}}^{2})
 + 2 \sqrt{2} S_{z}$ is,
\begin{eqnarray}
{\cal H}_{t}  &=&[{\cal H}_{1} + 2S(S+1) - 1]^{2}\nonumber \\
&=& 1 + 4({S_{z}}^{2} - {S_{x}}^{2})^{2} + 4[S(S+1) - {S_{y}}^{2}] 
\nonumber \\
&& \  \ \ \ \ \ \ + 8 \sqrt{2} ({S_{z}}^{2} - {S_{x}}^{2})S_{z},
\label{quinzebis}
\end{eqnarray}
which in the absence of the non-trivial Zeeman term $ 8 \sqrt{2} 
({S_{z}}^{2} - {S_{x}}^{2})S_{z}$ has $x \leftrightarrow z$ symmetry.  
In generalizing this approach to $n\ne 1$ this Zeeman term will change 
form with $n$.

The conclusion reached by all three methods is that the occurrence of 
simultaneous degeneracies with a four fold symmetry implies a 
non-trivial Zeeman term.  In nature molecular magnets can and will 
have such non-trivial time reversal symmetry breaking terms.  For 
example, when there is finite angular momentum in the crystal field 
ground state, and when this corresponds to a multiplet which is 
reflected by a fictitious spin, then it is entirely possible to have 
complicated Zeeman terms of the type which arise here.  Probably the 
best known example is the $\Gamma_{8}$ ground state of rare earth ions 
in cubic symmetry.

Although the simplest model with four fold symmetry, ${\cal H} = D 
{S_{z}}^{2} - a \left({S^{+}}^{4} + {S^{-}}^{4}\right) + g\mu_{B} B 
S_{z}$, does not have exact pseudo-Kramer's degeneracies, the present 
formalism {\it can\/} be used to determine {\it approximately\/} the 
fields at which the lowest {\it four\/} levels cross for not too small 
$S$.  For positive $D > a$, the models, Eqn.~(\ref{dixneuf}) and 
Eqn.~(\ref{dixneufbis}), yield different information.  With this sign 
of $D$ the diagonal anisotropy term is dominated by the ${S_{z}}^{2}$ 
term and this localizes the wave function near $S_{z} = 0$.  For model 
Eqn.~(\ref{dixneuf}), $2 S^{2}(\eta^{2}+\gamma^{2}) = D$ and the 
Zeeman term is in turn dominated by $4S^{2}(\eta^{2}-\gamma^{2}) S_{z} 
\approx 2D S_{z}$.  Directly, for model Eqn.~(\ref{dixneufbis}) the 
coefficient of ${S_{z}}^{2}$ is $D$ while the dominant part of the 
Zeeman term is only $D S_{z}$.  The difference in the two Zeeman terms 
reflects the different nature of the degeneracies as explained above.  
For integer spin $S$, the field such that $g \mu_{B} B = \hbar D$ 
deduced from Eqn.~(\ref{dixneufbis}) corresponds to the first 
half-integer point with two doublets while $g \mu_{B} B = 2\hbar D$ 
deduced from Eqn.~(\ref{dixneuf}) corresponds to the first finite 
field whole-integer point with a doublet and two singlets (and a 
singlet ground state).

The model Eqn.~(\ref{dixneufbis}) can also be used to estimate 
deviations from periodicity.  Again consider specifically positive 
$D>a$ and modestly large whole integer $S$.  The lowest lying states 
localize near $S_{z} \approx g \mu_{B} B /2\hbar D$ which implies an 
approximation: $Z({S_{z}}^{2}) \Rightarrow Z((g \mu_{B} B /2\hbar 
\hbar D)^{2})$.  Then e.g., for the low lying levels the first 
half-integer field $B_{1}$ is approximately determined by solving,
\begin{equation}
g \mu_{B} B_{1} = (D + \frac{a^{2}}{D}Z((g \mu_{B} B_{1} /2\hbar D)^{2})).
\label{last}
\end{equation}

It might also be noted that the anti-linear nature of $T$ is {\it 
not\/} essential to the development.  Consider some other 
``reflection'' symmetry $R$ such that $R^{2}=1$.  This must conserve 
the angular moment commutation rules so some components of $\vec S$ 
change sign while others do not.  For example, there exists a $R$ such 
that $RS_{x}R=-S_{x}$, $RS_{z}R=-S_{z}$ but $RS_{y}R=S_{y}$ for which 
$RrR = - r^{\dagger}$ and for which the above development is 
essentially unchanged.

In the much used functional integral method degeneracies correspond to 
the destructive interference of two equivalent tunneling paths and 
this implies that the topological phases are $\pm \pi/2$.  This can 
also be verified using the current analytic techniques.  Consider 
again the model Eqn.~(\ref{six}), the topological phase can be 
``measured'' as a Berry phase\cite{berry} for a suitably conceived 
adiabatic process.  The two states $\left| n, \alpha \right\rangle$, 
$\alpha = e$ or $o$ have $\langle S_{x} \rangle =0$.  The imagined 
adiabatic process is therefore between the states $\left| n, \pm 
\right\rangle = (2)^{-1/2}[ \left| n,e \right\rangle \pm \left|n, o 
\right\rangle ]$ which have $\langle S_{x} \rangle \approx \pm S$.  
The tunneling path which passes via the positive or negative $y$-axis 
is favored by adding $\mp h_{t}S_{y}$, $h_{t} = g \mu_{B} B_{y} > 0$.  
Observe first that with $h_{t} = 0$ the matrix ${\cal H}_{n}$ is real 
and therefore solutions $\left|n , \pm \right\rangle$ can be made 
real.  On the other hand $S_{y}$ is a pure imaginary matrix.  It is 
trivial that there are pure imaginary matrix elements proportional to 
$h_{t}$ which connect the $\left| n,\pm \right\rangle$.  With a very 
small $h_{t}$ the magnetization will therefore slowly and 
approximately adiabatically oscillate between $S_{x} \approx \pm S$ 
with one of the above least action paths dominating.  Reflecting the 
pure imaginary matrix elements, if the initial state is $\left|n,+ 
\right\rangle$ then after a half-period the system will find itself in 
the state $ - i {\rm sign}(h_{t}) \left|n,- \right\rangle$ 
demonstrating that the Berry phase is $\mp \pi/2$.

Especially in one dimension, the nature of the ground state of a 
concentrated magnet depends on the half or whole-integer nature of the 
spins with whole integers having gaps\cite{gap}.  It is possible to 
imagine, e.g., a magnet with Eqn.~(\ref{un}) describing the 
intra-atomic interactions and with, e.g., $-\sum_{ij}J_{ij}\vec 
S_{i}\cdot \vec S_{j}$ interatomic interactions.  If the spin $S$ is 
whole integer, and if $J$ is suitably small compared with the 
intra-atomic splittings $\sim D$, then a field with one of the magic 
values $B_{n}$ will transmute the system to one with effective 
half-integer spin $S-(2n-1)/2$.  Since the $B_{n}$ values are periodic 
the phase diagram will also be periodic in the applied field.

The period measured in terms of the magnetization per site can be 
determined assuming that the magnetization is given to a good 
approximation by the classical result.  (Even for a small spin value, 
say $S=2$, this is a good approximation for most parameters ranges.)  
The term $-(D+E) {S_{x}}^{2}$ might be considered to be a $x$-directed 
anisotropy field $B_{A}$ such that $g\mu_{B} B_{A} = 2S\hbar (D+E)$.  
In the presence of a real field $B$ along the $z$-direction the 
dimensionless magnetization component along this direction is $m 
\approx S \tan (h/2S\hbar (D+E)) \approx (h/2\hbar (D+E))$ and the 
period of the whole integer points is $\Delta m = \sqrt{E/(D+E)} \le 
1$.  Thus magnetization plateaus associated with the gaped whole 
integer phases will {\it not\/} occur at simple multiples of $g 
\mu_{B}$ as is often supposed to be the case.

\end{document}